# Highly efficient collective coupling between laser diode array and lensed fibre ribbon

P. Chanclou, M. Thual, J. Lostec, P. Auvray,
J. Caulet, G. Joulié, A. Poudoulec, and B. Clavel

FRANCE TELECOM, CNET, DTD/PIH
2, avenue Pierre Marzin, F22307 LANNION CEDEX FRANCE

## ABSTRACT

A new concept is proposed for lensed fibres fabricated according to a collective and low cost process. This process is based on the cleaving and splicing of optical fibre ribbons and is suitable for the coupling of laser diode arrays and fibre ribbons.

**keywords**: optical coupling, laser array, lenses, laser beam focusing, optical fibres ribbon.

## 1. INTRODUCTION

The implementation of optical communications in the access network involves reducing the cost of optical modules. The assembly step of these modules remains the most important part of this cost. Regarding transceiver modules in the central office, a way to lower the packaging cost could be to apply a collective and passive alignment process ( without activating the laser array during alignment ). To achieve such a collective assembly, we need collective coupling micro-optics. The characteristics of this micro-optics should be compatible with the assembly tolerances and should allow achieving a satisfactory coupling[1]. The main features[2] of this coupling arrangement are high efficiency coupling, relaxed alignment tolerances, long working distance, small packaging volume and low cost fabrication process. The aim of this work is to design a collective process to fabricate micro-optics at the end face of ribbon fibres that fulfil these requirements.

## 2. PRINCIPLE AND FABRICATION

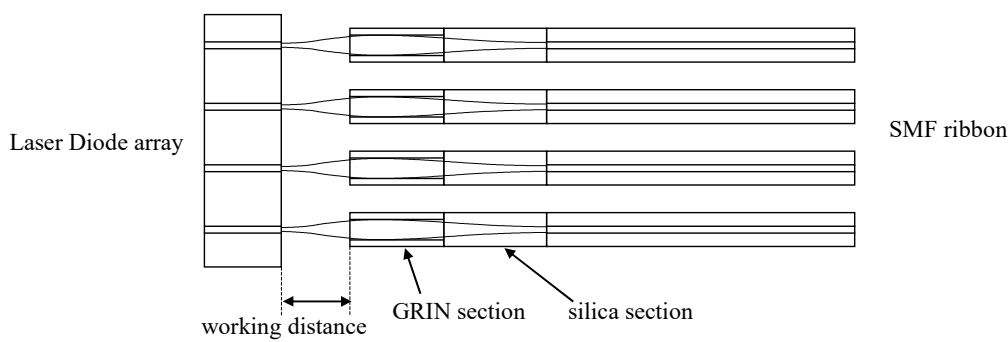

Fig. 1. Schematic representation of optical coupling arrangement between Laser Diode array and SMF ribbon.

In the present paper, a new concept[3], proposed for a lensed fibre ribbon, requires a simple collective fabrication process. As shown in Fig. 1, this optics consists of a graded index (GRIN) fibre section which acts as a GRIN lens[4]. This latter is maintained at the proper distance from the 1300/1500 nm single-mode fibre (SMF) by a coreless silica fibre section. The micro-optics are made with standard ribbons composed of four 250 μm spaced fibres. We used commercially available fibre : G652 single-mode fibre and 85/125/250 Alcatel graded index fibre. This multimode fibre used has a core diameter of 85 μm, a refractive index on axis equal to 1.472 and a quadratic gradient constant $g = 4.326$ mm$^{-1}$.

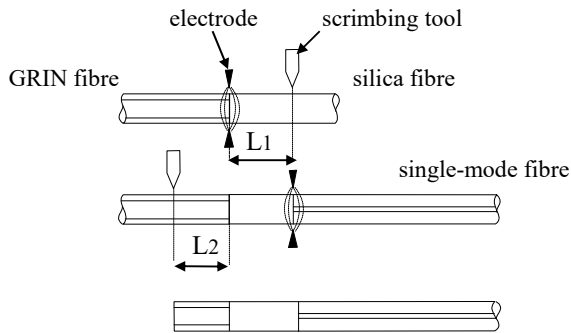

Fig. 2. Schematic illustration of the micro-optics fabrication process.

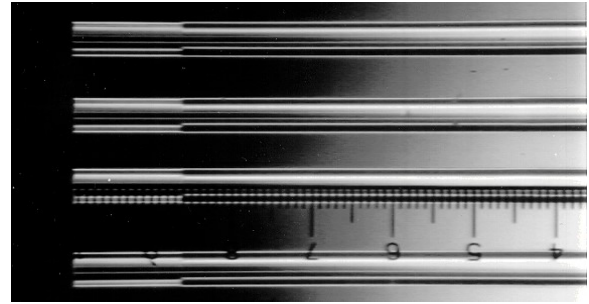

Fig. 3. Photograph of the micro-optics ribbon.

The fabrication process is illustrated in Fig. 2. Using an arc-discharge fibre splicer, the GRIN ribbon is first spliced to the silica ribbon. The silica ribbon is then cut at a distance $L_1$ from the silica/GRIN splicing. The GRIN and silica ribbon are spliced to the SMF ribbon. Finally the GRIN part of ribbon is cut at a distance $L_2$ from the silica/GRIN splicing. Fig. 3 shows a view of the micro-optics ribbon. The reproducibility of the lengths is about ±2 μm for each section of ribbon.

This micro-optics is composed of two splices with two different fibres. In order to examine the optical interfaces of these splices, the micro-optics has been burnished in a cross-section. The figure 4 shows a picture of the micro-optics in a half cross-section by a scanning electron microscope (SME).

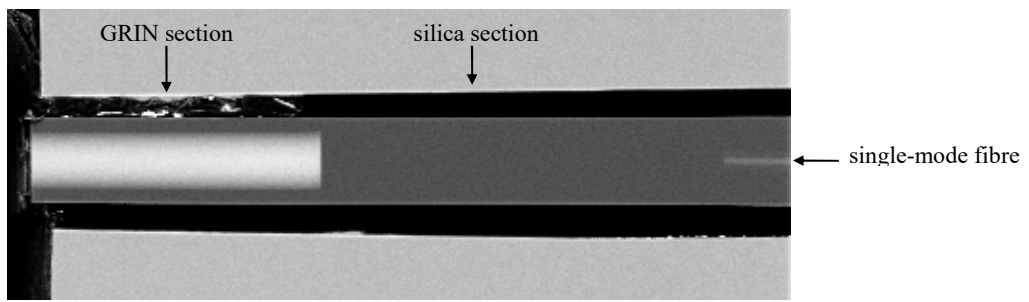

Fig. 4. SEM picture of the micro-optics.

As shown by the picture, the optical interfaces between different fibres are well defined. Both splice losses have been measured below 0.1 dB. The fibre outer diameter of 125 μm remains unchanged along the lensed fibres. Note that this collective micro-optics ribbon only requires a collective fibre splicer and cleaver.

### 3. MICRO-OPTICS PROPERTIES

As an example the mode field diameter was measured at the micro-optics focal point. This focal point may be either true, or virtual. Near field and far field experiments were carried out ( see Fig. 5 ). Length $L_1$ of the silica section is fixed at 0 and 400 μm, as shown in Fig. 5 (a) and (b) respectively. Length $L_2$ of the GRIN section is extended to modify the focal length of the GRIN lens. The first configuration with $L_1 = 0$ μm is a selfoc configuration.

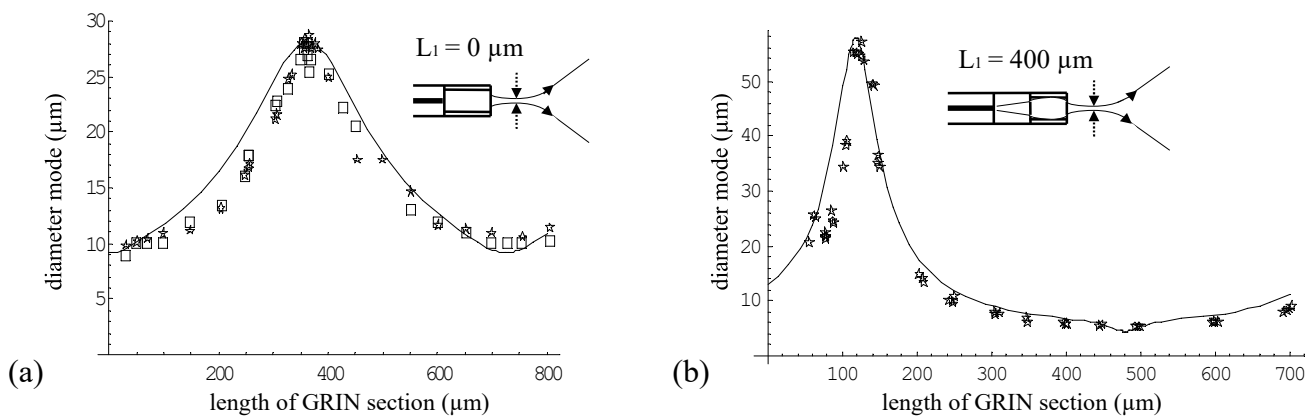

Fig. 5. Far field ( box ) and near field ( star ) of micro-optics with a length of silica $L_1 = 0$ µm for (a) and $L_1 = 400$ µm for (b). Continuous lines are theoretical curves.

As shown in Fig. 5, experimental results are in good agreement with theoretical curves. Compared to the selfoc configuration ( $L_1 = 0$ µm ) the micro-optics ( $L_1 = 400$ µm ) provides a very wide range of mode diameters at the focal point. The mode diameter range obtained by the micro-optics is 6 to 60 µm instead of 9 to 28 µm for the selfoc. This is due to the fact that the silica section extends the mode field at the beginning of the GRIN section.

The distance between the end face of the micro-optics and the true focal point is the working distance. In the case of the selfoc, the working distance is limited in the 0 to 230 µm range while it is extended up to 900 µm when using micro-optics.

The lengths of the silica and GRIN sections are defined to optimise the coupling efficiency between the LD array and SMF ribbon. Note that this optimising takes into account the fact that the beam propagation is truncated by the optical cladding of the GRIN sections. The optimum lengths of silica and GRIN sections for the LD coupling is a compromise between the equality of mode diameters and the truncation of beam ( not detailed in this paper ). The lengths could also be optimised for other applications that provides a good flexibility.

## 3. OPTICAL COUPLING CHARACTERISTICS

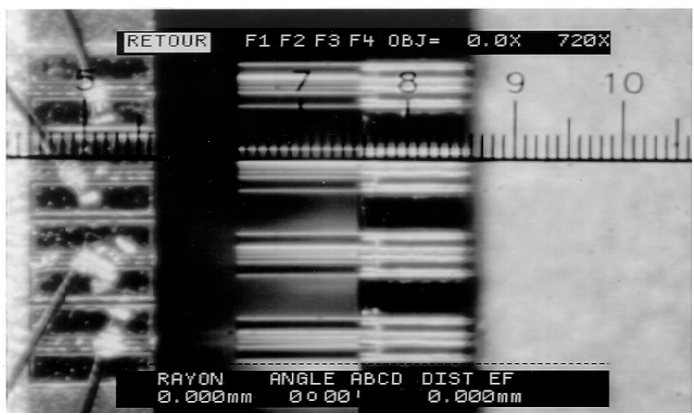

Fig. 6. Photograph of four-channel LD array
chip coupled with a micro-optic ribbon.

The micro-optics ribbon is held in a silicon fibre-carrier fitted with four 250 µm spaced V-grooves. For these experiments we used two types of Laser Diode having an operating wavelength of 1.3 µm. The first one is a conventional buried ridge stripe LD ( Alcatel Optronics ). It has emitting half angles (and radius mode) defined by the far-field at $1/e^2$ of the maximum intensity of 25° ($\omega = 0.89$ µm) and 30° ($\omega = 0.71$ µm) in the parallel and perpendicular directions against the epitaxial surface, respectively. The second one is a tapered LD ( Mitsubishi ) whose emitting half angles (and radius mode) are 10.1° (ω = 2.37 μm) and 12.6° (ω = 1.98 μm) in the parallel and perpendicular directions. Fig. 4 shows the lensed ribbon coupled with the LD array.
In order to evaluate the homogeneity of the micro-optics ( Fig. 7 ), the coupling efficiency of each fibre of the ribbon is measured at the optimum position in front of the same LD. The achieved coupling efficiency, defined as 10log(Pcoupled/Pemitted LD), is -4.54±0.04 dB for the conventional LD with a working distance of 48±0.8 μm and -1.45±0.06 dB for the tapered LD with a working distance of 50±0.9 μm.

(a)

| conventional LD | Coupling Loss dB | Working distance μm |
|---|---|---|
| fibre 1 | -4,5 | 48,7 |
| fibre 2 | -4,58 | 48,2 |
| fibre 3 | -4,55 | 47,6 |
| fibre 4 | -4,54 | 47,2 |

(b)

| tapered LD | Coupling Loss dB | Working distance μm |
|---|---|---|
| fibre 1 | -1,51 | 50,9 |
| fibre 2 | -1,39 | 49,5 |
| fibre 3 | -1,45 | 49,4 |
| fibre 4 | -1,44 | 50,5 |

Fig. 7. Optical coupling efficiency of each fibre of the ribbon measured at the optimum position in front of the same (a) conventional and (b) tapered LD.

The theoretical coupling loss calculated by the beam propagation method, or by the gaussian beam propagation method[5,6] is -4.5 dB ( not detailed in this paper ).
   The measurements of the coupling loss as a function of axial and lateral displacements for both LDs, are shown in Fig. 8 and Fig. 9 The optical coupling of the micro optics is compared with a butt coupling of cleaved 1300/1500 nm SMF. Loss increments due to displacements in the x and y directions are almost the same, hence only one curve was plotted.

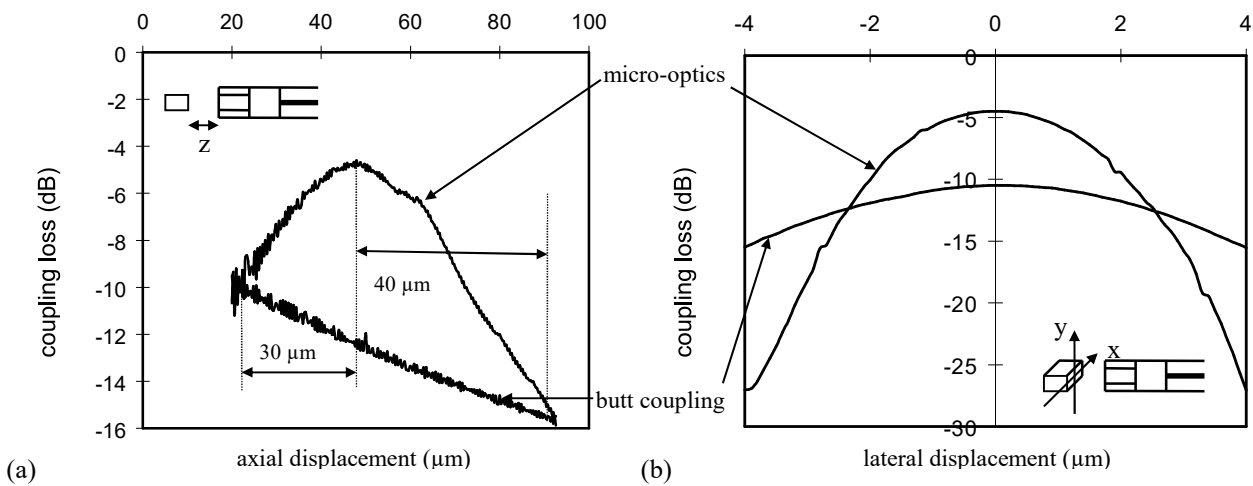

Fig. 8. Coupling-loss characteristics as a function of axial displacement (a) and lateral displacement (b) in front of conventional LD.

When the micro-optics is coupled to a conventional LD ( Fig. 8 ), the optical alignment tolerance along the axis is ±9 μm for 1 dB excess loss. The micro-optics is better than SMF in the range between +30 μm and -40 μm around the optimum coupling position of micro-optics for axial displacement. The lateral tolerance is ±1 μm for 1 dB excess loss. For this lateral displacement the micro-optics is better than SMF in the ±2.4 μm range.

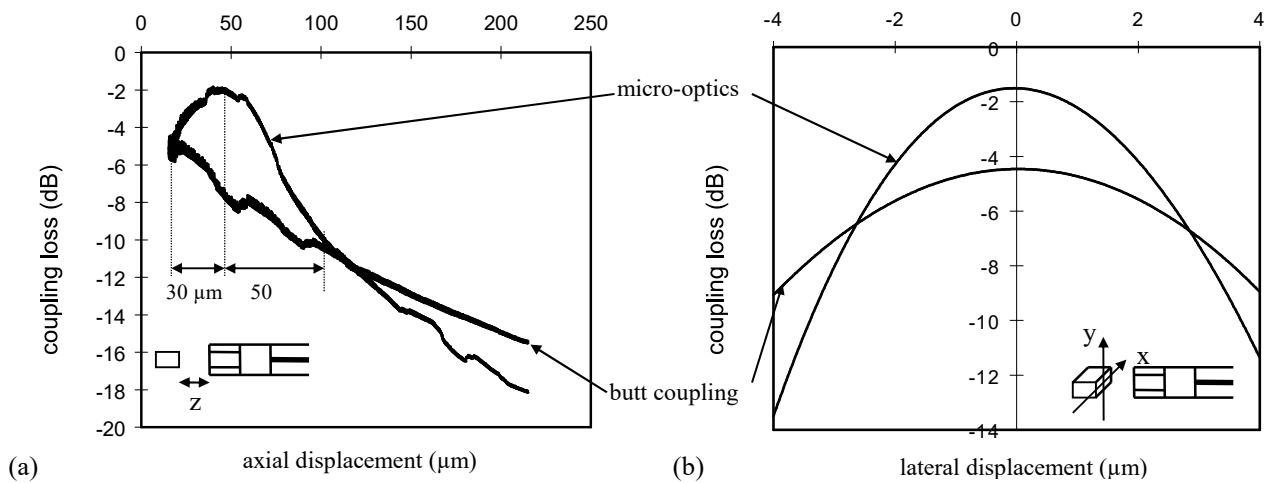

Fig. 9. Coupling-loss characteristics as a function of axial displacement (a)
and lateral displacement (b) in front of tapered LD.

When the micro-optics is coupled to a tapered LD ( Fig. 9 ), the optical alignment tolerance is ±15 µm in the axial direction for 1 dB excess loss. The micro-optics is better than SMF in the range between +30 µm and -50 µm around the optimum coupling position of micro-optics for axial displacement. The lateral tolerance is ±1.2 µm for 1 dB excess loss. For this lateral displacement the micro-optics is better than SMF in the ±2.7 µm range.

Fig. 10 shows a comparison between the results of coupling loss of micro-optics in front of conventional LD and tapered LD. It may be seen that using tapered LD decreases the coupling loss and slightly improves the lateral and axial tolerances.

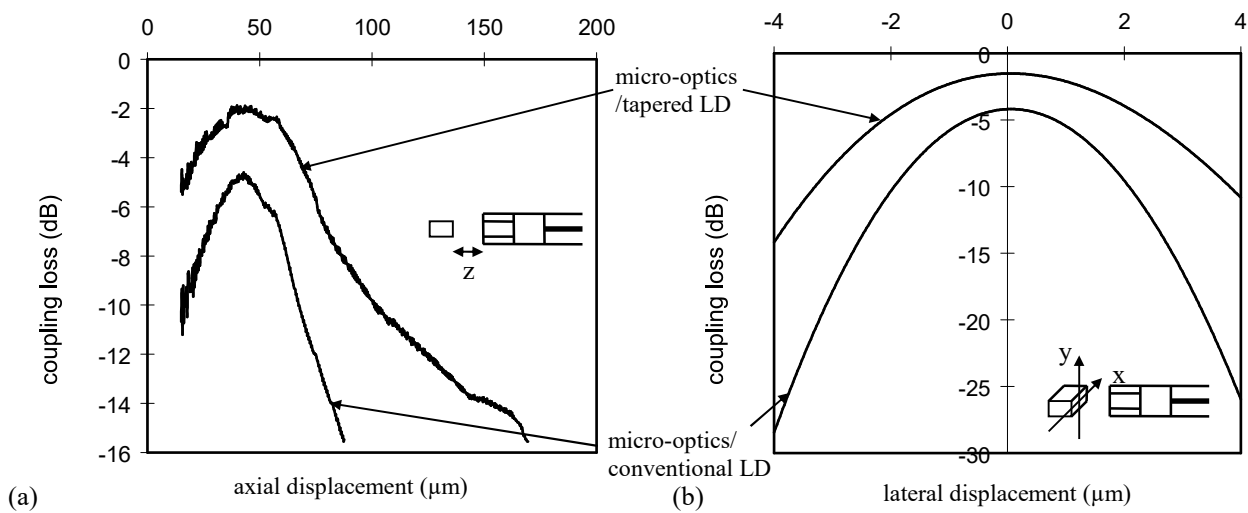

Fig. 10. Coupling loss characteristics of the micro-optics in front of conventional LD and tapered LD as a function of axial displacement (a) and lateral displacement (b).

Coupling a tapered LD with this micro-optics is the most suitable arrangement for optimising coupling efficiency and optical alignment tolerances.

The length tolerances of the GRIN and silica sections are ±10 µm and ±50 µm respectively, for 1 dB excess loss of the coupling loss between a conventional LD and the micro-optics. The length tolerances are wider than the cleaving precision of ±2 µm. This micro-optics has a tolerant fabrication process.

## 5. CONCLUSION

Our arrangement provides several attractive features in comparison with conventional coupling techniques. The lensed fibre ribbon is fabricated using a collective and simple process which allows reproducibility and mass-production. Long working distances, relatively high coupling efficiency and a considerable positioning tolerance have been obtained. This micro-optics is in full agreement with the specification of working distance and optical alignment tolerances required for packaging technique. This new concept is promising for production of multichannel optoelectronic modules.

## 6. ACKNOWLEDGEMENT

The authors would like to thank N. Devoldère and D. Pavy for valuable comments on this work, P. Grosso for providing silica fibres, J. C. Bizeul for fibres characterisations, C. Vassalo for fruitful theoretical discussions, D. Rivière for SEM pictures, and G. Audibert and J. Salaün for mechanical support.

— Further author information —
Email: philippe.chanclou@cnet.francetelecom.fr; WWW: http://cnet.fr; Telephone: 331 96 05 35 81; Fax: 331 95 05 20 65
Email: monique.thual@cnet.francetelecom.fr; WWW: http://cnet.fr; Telephone: 331 96 05 20 28; Fax: 331 95 05 20 65